\documentstyle[twoside,fleqn,espcrc2,epsfig]{article}

\newcommand{\AmS}{{\protect\the\textfont2
  A\kern-.1667em\lower.5ex\hbox{M}\kern-.125emS}}

\centerline {\bf ~~~~~~~~~~Review of Cosmic Ray experiments with underground detectors}
\centerline {\bf ~~~~~~~~~~~~~~~~~E. Scapparone~~~~~~~~~}
\centerline{~~~~~~~~~~INFN- Laboratori Nazionali del Gran Sasso}
\centerline{~~~~~~~~~~ S.S.17 km 18+910, 61070, Assergi (AQ), Italy}      
\centerline{ \bf ~~~~~~~~~~Invited talk at 6th International Conference on advanced }
\centerline{ \bf ~~~~~~~~~~Technology and Particle Physics}
\centerline{\bf ~~~~~~~~~~Villa Olmo, Como, Italy, October 5-9, 1998}
\centerline{\bf ~~~~~~~~~~To be published in Nucl. Phys. B, Proc. Suppl.}
\title{Review of Cosmic Ray experiments with underground detectors}
\author{E. Scapparone\address{INFN, LNGS
        S.S.17 km 18+910, 61070, Assergi (AQ), Italy}}
\begin{document}

\begin{abstract}
The most important underground detectors addressing Cosmic Ray physics 
are described, with a special emphasis on the description
of the used technology.
\end{abstract}\vspace{0.2cm}
\maketitle
\section{Introduction}
The first generation of underground experiments ran during the '80s,
triggered by the GUT theories. The most important
results they obtained were the IMB analysis on proton decay, ruling
out the minimal SU(5) model, the $~^{8}$B solar neutrino observation by
Kamiokande and the Supernova SN1987 detection.
The impact on the technology development was relevant. Let me quote for
instance the NUSEX experiment\cite{NUSEX}: this detector
proved the reliability of digital calorimeters based on plastic streamer
tubes, opening the road to the construction of the hadronic calorimeters
for the LEP experiments(ALEPH\cite{aleph}, OPAL\cite{opal}). 
IMB and Kamiokande showed that the water Cherenkov technique was robust while
LSD pointed out the possibility of using liquid scintillator to search for Supernova
explosion. The main limitation of these detectors was the 
collected statistics . The natural development was therefore the construction 
of similar detectors
enlarged by an order of magnitude in volume or in area.
The experience gained with NUSEX allowed the construction of MACRO\cite{macro},
with an acceptance increased to more than two orders of magnitude. The Kamiokande 
and the IMB Collaborations joined Super-Kamiokande, reaching a mass of 50 Ktons and LSD inspired LVD at Gran Sasso, actually running with 600 tons of liquid
scintillator.
The most important questions are: how much the technology of these 
experiments is advanced ?
Do the most important technological limitations come from the large 
volume/area used or this kind of physics requires just a modest detector 
performance ? Could they benefit of more sophisticated techniques ? 
I will try to answer to these questions reviewing the main 
underground detectors for Cosmic Ray physics actually running.
\section{MACRO}
Let me start with
the MACRO experiment, located in the HALL B of Laboratori Nazionali del Gran
Sasso. This is a multi purpose experiment for GUT magnetic monopole search, 
atmospheric and Supernova neutrino and Cosmic Ray studies. The detector 
consists
of 14 horizontal layers of plastic streamer tubes(PST) and 3 horizontal layers
of liquid scintillator(LS). Each side is closed with a sandwich of 6 PST
and a layer of LS. 
The cell size of the streamer tube system has a cross section of 3$\times$3
c$m^{2}$, larger than the NUSEX one (1$\times$1 c$m^{2}$). The readout
was performed in NUSEX by using orthogonal pick-up strips, while in MACRO 
the wires and a layer of pick-up strips, forming an angle of 26.$5^{o}$
with the direction orthogonal to the wires, are read. The larger cell size,
due to the dominant effect of the multiple scattering of the muons through
the rock, doesn't affect significantly the pointing accuracy.
Fig.1 shows the angle between multiple muons detected by 
MACRO. Since high energy multiple muons can be considered parallel within
few milliradians, the distribution is the folding of the muon 
multiple scattering with the detector angular resolution. 
\begin{figure}[htb]
\vskip .5 cm
\begin{center}
\vskip -1cm
\mbox{\epsfysize=75mm 
      \epsffile{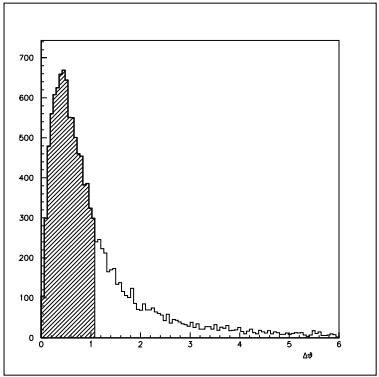}}
\end{center}
\vskip -.5 cm
\caption{\em  Space angle $\Delta$$\theta$(degrees) 
between the direction of tracks in two muon
events. The dashed area contains the 68$\%$ of the distribution.}
\label{f:streamer}
\vskip -0.8cm
\end{figure}
The 68$\%$ of the events has an angle $\theta$$\leq$1.$1^{o}$ and 
therefore $\sigma_{\theta}$=1.$1^{o}$/$\sqrt{2}$$\simeq$0.$8^{o}$. The 
distribution is largely dominated by the multiple scattering,
whose contribution at this depth is around $\sigma_{\theta}^{MS}$$\simeq$
0.$6^{o}$. This is a typical example in which the granularity of the 
detector is limited by an external source of indetermination; an 
improvement of the detector space resolution would result just in a small
improvement of the muon position measurement and pointing accuracy.
\begin{figure}[htb]
\vskip -1.2 cm
\begin{center}
\mbox{\epsfysize=75mm 
      \epsffile{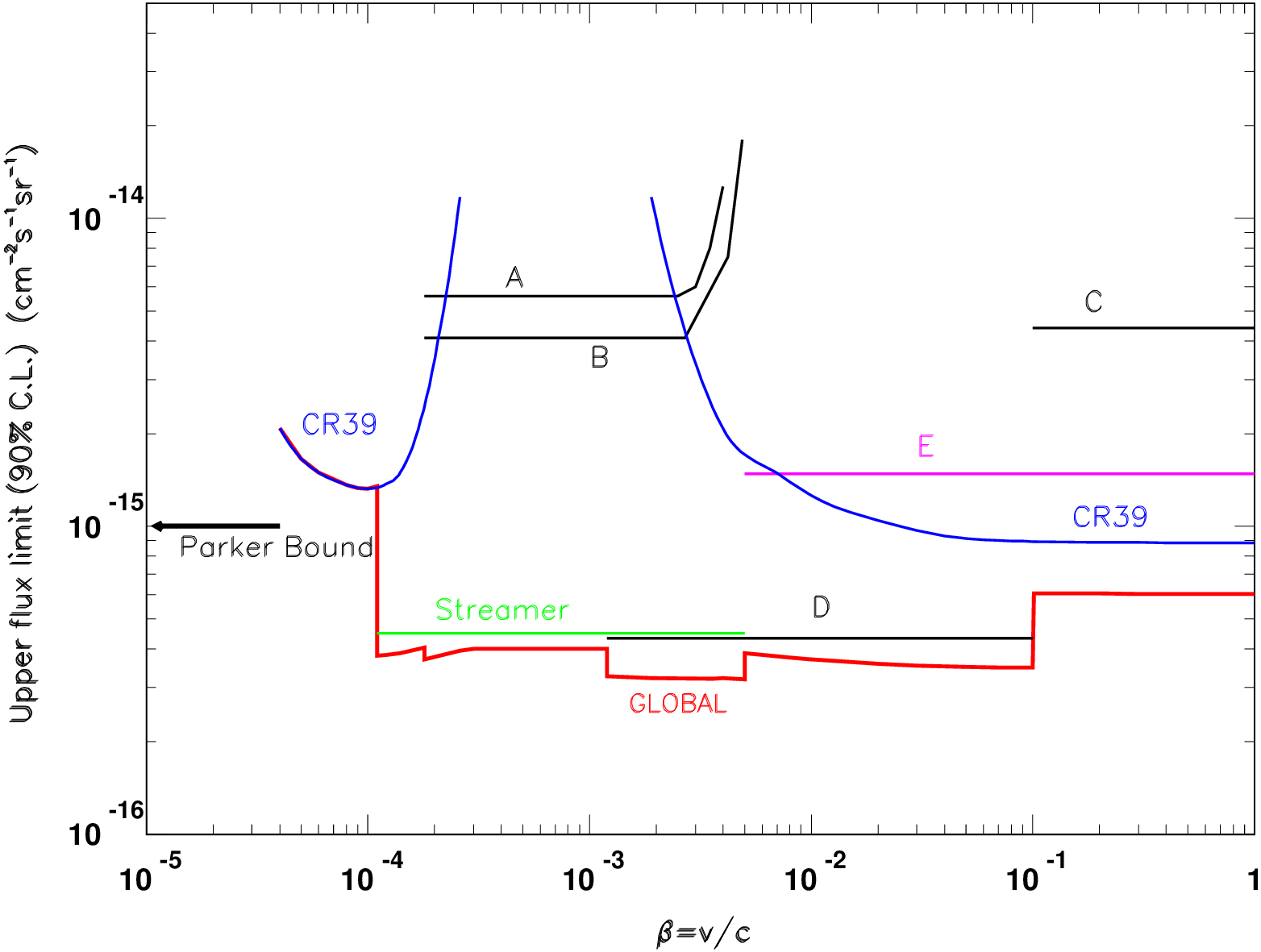}}
\end{center}
\vskip -2.1 cm
\caption{\em MACRO monopole flux upper limit.}
\label{monopole}
\vskip -0.8cm
\end{figure}
Fig. 2\cite{monocei} shows the MACRO magnetic monopole flux upper limit.
Different techniques are used to identify a monopole in the various
$\beta$ regions explored. At (1$0^{-4}$$\leq$$\beta$$\leq$1$0^{-3})$ the 
liquid scintillator informations are very helpful:
the magnetic monopole signature is a train of photoelectrons, with a 
duration proportional to the monopole velocity. 
\begin{figure}[tb]
\begin{center}
\vskip 2cm
\mbox{\epsfysize=75mm 
      \epsffile{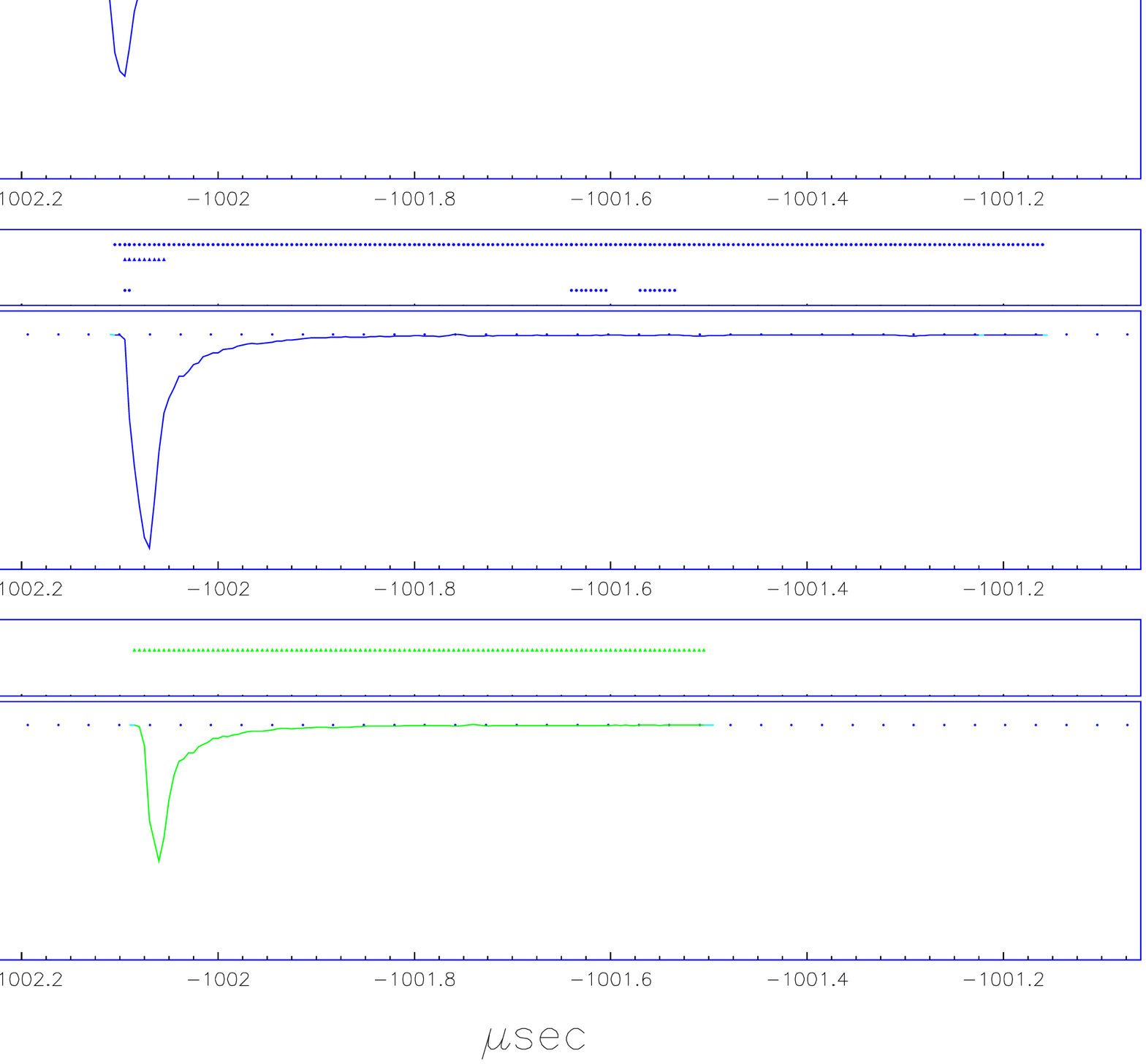}}
\end{center}
\vskip -3.5 cm
\caption{\em Muon signal in the MACRO scintillator obtained with the 
200 MHz Waveform digitizer \label{f:wave}.}
\vskip -1.cm
\end{figure}
This is a delicate search,
since background contamination could mimic the signal: a simple ADC/TDC system
is not enough! MACRO takes advantage of a Waveform Digitizer System, with
a sampling frequency of 200 MHz. The system is calibrated using LED light.
Fig. 3 \cite{tesiivan}shows a muon crossing three layers of liquid scintillator and
the respective Waveforms.
The study of upgoing muons, originating from neutrino interaction below 
the apparatus, requires high detector performance too. 
The discrimination against
the much more abundant downgoing muons is obtained by using the time 
of flight system (TOF). The rejection factor required for this analysis is
$\simeq$1$0^{-4}$. Fig. 4 shows the 1$/$$\beta$ 
distribution obtained by MACRO. The left peak 
comes from upgoing muons, while the right peak comes from 
downgoing muons. The TOF resolution is better than 1ns, allowing a clear
separation between the two peaks. Of course higher timing performance 
with TOF systems have been reached, for instance a resolution 
$\sigma_{t}$$\simeq$100 ps has been
obtained by the AMS TOF system\cite{ams}. Nevertheless, it must be stressed 
that in this case 
a $\sigma_{t}$$\leq$1ns resolution is obtained on a huge scintillator mass 
(600 tons), making use of 12 meters length counters,
running for a long period, more than 8 years.    
As far as the physics results are concerned, such TOF system performance 
allowed the study of atmospheric neutrino anomaly. MACRO pointed out
\cite{atmmacro} a 
discrepancy between the real data and the Monte Carlo prediction for 
atmospheric $\nu_{\mu}$ flux . 
The ratio between the data and the Monte Carlo prediction
for upgoing muon flux was R=0.74$\pm$0.036(stat)$\pm$0.046(sys)$\pm$
0.13(theoretical).
An improvement of the agreement between Monte Carlo and the MACRO data is found
supposing an oscillation 
$\nu_\mu$$\rightarrow$$\nu_\tau$ or $\nu_\mu$$\rightarrow$$\nu_s$ with
$\Delta$$m^{2}$=2.5$\cdot$1$0^{-3}$e$V^{2}$ and si$n^{2}$=1.0.
Further informations about neutrino oscillation could come from the 
measurement of the muon energy.
Such a measurement for through-going muons is very difficult with 
the present generation of underground experiment. This is a typical situation
in which a more advanced technology would have been useful. 
The average energy of  upgoing muon is $\simeq$ 20 GeV, while the 
average energy of
downgoing muons at LNGS rock depth is $E_{\mu}$$\simeq$300 GeV. 
A first attempt to measure the muon energy is due to the 
NUSEX and to the LVD Collaborations.
\begin{figure}[tb]
\vskip -2.2 cm
\begin{center}
\mbox{\epsfysize=75mm 
      \epsffile{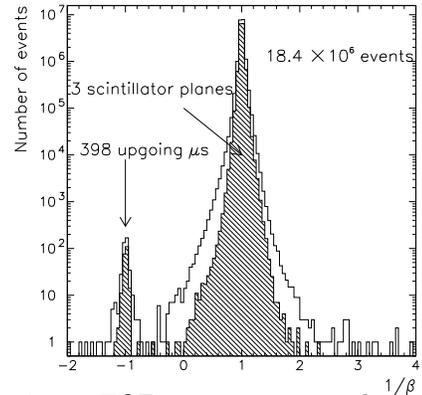}}
\end{center}
\vskip -1.8 cm
\caption{\em TOF measurement obtained by MACRO: right peak downgoing
muons, left peak upgoing muons\label{f:tof}.}
\vskip -0.8cm
\end{figure}
\section {LVD}
LVD is a multipurpose detector consisting of a large volume of liquid 
scintillator interleaved with limited streamer tubes in a compact geometry
\cite{lvd}. The apparatus has a modulare structure that consists of aligned
towers of 38 modules each. Every module contains 8 liquid scintillator 
counters of dimension 1.5$\times$1$\times$1 $m^{3}$ seen by three 
photomultipliers. Taking advantage of the L shaped detectors used as 
tracking system, LVD is able to measure the depth-intensity curve up to
20 Kmwe( Kilometers of water equivalent). As far as the measurement of
muon energy is concerned, LVD measured the quantity 
$<$$\Delta$E/$\Delta$L$>$ as a function of the rock depth h 
, where L is the track length in the scintillator counter and 
$\Delta$E is the energy lost in the counters.
For h$\leq$8 Kmwe, this observable
is expected to increase with h, since a selection of higher rock
depth corresponds to higher threshold for Cosmic Ray muon energy.
For h$>$8 Kmwe, the muon flux coming from atmospheric muon becomes
negligible. Observed muons come from atmospheric neutrino interaction in 
the rock in the neighborhood of the detector: 
$\nu_{\mu}$+N$\rightarrow$$\mu$+X.
The different origin of the muons, manifests for instance in the change
of their average energy. This is small signal, difficult to observe.
Fig. 5 shows 
$<$$\Delta$E/$\Delta$L$>$ as a function of h. Although the statical error
for high depth is large, at h$>$8 Kmwe a clear decrease of 
$<$$\Delta$E/$\Delta$L$>$ is visible in the LVD data.
\begin{figure}[tb]
\begin{center}
\vskip -1.5 cm
\mbox{\epsfysize=75mm 
      \epsffile{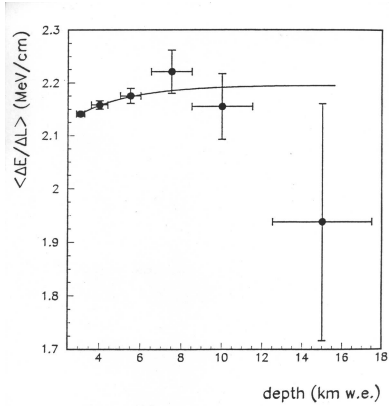}}
\end{center}
\vskip -1.2 cm
\caption{\em Measurement of $<$$\Delta$E/$\Delta$L$>$ 
in the LVD scintillator \label{f:lvdenergy}as a function of the slant depth.}
\vskip -1.cm
\end{figure}
\section {Precision measurements}
Cosmic ray detectors are usually supposed to perform measurements only 
with 10-20$\%$ accuracy. I would like to show two examples of small 
signals observed deep underground. Let me start with the measurement 
of the distribution of the distance between muon pairs detected deep underground,
the so called ''decoherence''. This distribution depends on different cosmic
rays features: the C.R. cross section, the muon parent mesons $p_{t}$
distribution, the multiple scattering of the muons through the rock. 
 A special care has been devoted to the study of the tail at high distance
of this distribution, to search for
anomalous high $P_{t}$. On the contrary the surprise came
from the low distance region. Fig. 6 shows the decoherence
function observed by MACRO. A good agreement is evident between the MACRO 
data and the HEMAS Monte Carlo expectation. 
\begin{figure}[tb]
\begin{center}
\vskip 1.cm
\mbox{\epsfysize=75mm 
      \epsffile{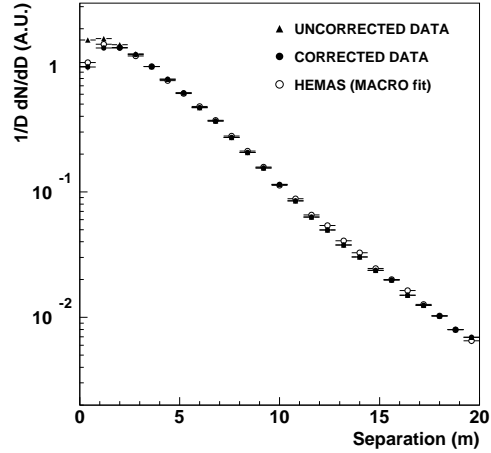}}
\vskip -3. cm
\caption{\em The low distance region of the MACRO experimental
decoherence before and after the subtraction of the 
$\mu$+N$\rightarrow$$\mu$+N+$\mu^{+}$+$\mu^{-}$ process.
\label{f:deco_clean}}
\end{center}
\vskip -1.cm
\end{figure}
Nevertheless, a discrepancy 
of 34$\%$ is found for muon separation D$\leq$80 cm. The same effect has been
pointed out
by the LVD Collaboration\cite{olga}: again the real data shows an excess
respect to the Monte Carlo. 
 An explanation of this effect has been proposed by the LVD Collaboration:
muons at short distance can come from the direct muon pairs production 
by muons: $\mu$+N$\rightarrow$$\mu$+N+$\mu^{+}$+$\mu^{-}$. It has
been pointed out\cite{kelner} that 
such process is suppressed, respect to the electron pair production, by a 
factor $m_{e}^{2}/m_{\mu}^{2}$ only for small v=$E_{\gamma}$/$E_{\mu}$. 
At high v the cross section is not suppressed by such factor. Since the 
produced muon pair lies at low distance from the original muon, the results
is an excess of events at short distance in the decoherence function.
A data analysis from MACRO\cite{macrodeco} shows that after the inclusion of this process
in the muon transport simulation code, the agreement between data and 
Monte Carlo is restored.
Another example of small signal pointed out in cosmic rays physics 
is the detection of photonuclear
interaction of muons deep underground. MACRO looked for charged hadrons
with E$\geq$300 MeV,
produced in the rock surrounding the detector\cite{macrodeco}.
A data analysis based on 
11,000 hours of live time, they found 
1,938 candidate events over a total sample of
9,544,318 muon events.
The result was expressed in terms of the ratio 
$R_{\mu+h}$ of the selected $\mu$+hadrons events 
to the number of muon events in the same live time.
After the subtraction of the background, they found for $R_{\mu+h}$ 
in real data and in the MC simulations :\\
\noindent- $R_{\mu+h}^{DATA}$~~~$=(1.91\pm0.05_{sta}\pm0.03_{sys})\cdot10^{-4}$,\\
- $R_{\mu+h}^{FLUKA}=(1.89\pm0.16_{sta}\pm0.02_{sys})\cdot10^{-4}$,\\ 
- $R_{\mu+h}^{GEANT}=(1.31\pm0.14_{sta}\pm0.02_{sys})\cdot10^{-5}$, \\
confirming thus the obsolete treatment of the muon photonuclear interaction
in GEANT3.15-3.21\cite{geant} pointed out in\cite{battistoni}.
\section{Super-Kamiokande}
\begin{figure}[tb]
\vskip -2. cm
\begin{center}
\vskip -0.7cm
\mbox{\epsfysize=75mm\epsffile{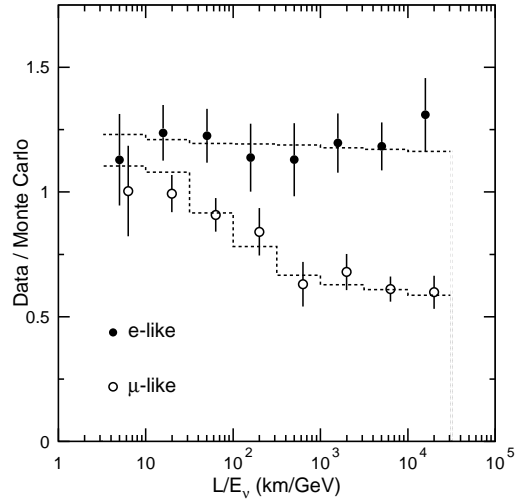}}
\end{center}
\vskip -1.2cm
\caption{\em ($\Phi_{\nu_{\mu}}$$)_{data}$/($\Phi_{\nu_{\mu}}$$)_{MC}$ as 
a function of L/E obtained by SuperK\label{f:sk}.}
\vskip -.8cm
\end{figure}

The Super-Kamiokande experiment is the underground detector with the
largest mass actually running. Events produced in 50 Ktons of water are 
reconstructed by 11,146 PMTs. 
The main goals of the experiment are 
the study of $~^{8}$B solar neutrino, the search for proton decay and 
the study of atmospheric neutrino. 
As far as the atmospheric neutrino is concerned, several measurements
can be performed by SuperK: the $\nu_{\mu}$/$\nu_{e}$ ratio,
the measurement of stopping/through-going muons and the measurement of the
$\nu_{\mu}$ flux as a function of L/E, where L is the distance travelled
by neutrino and E is the neutrino
energy.
The analysis of the ratio R=($\nu_{\mu}$/$\nu_{e}$$)_{data}$/
($\nu_{\mu}$/$\nu_{e}$$)_{MC}$ has been performed by several 
experiments. The NUSEX measurement\cite{atmnusex} was affected by a large
statistical error: R=(0.96$^{+.32}_{-.28}$), while the FREJUS Collaboration
reported a substantial agreement between data and Monte Carlo\cite{frejus}:
R=(1.00$\pm$0.15(stat)$\pm$0.08(sys)). On the contrary a deficit of atmospheric
$\nu_{\mu}$ has been measured by two Cherenkov detectors. IMB\cite{imb} 
found a ratio R=(0.54$\pm$0.05$\pm$0.11) and Kamiokande\cite{kamiokande}
measured R=(0.60$\pm$0.05$\pm$0.05).
The apparent mismatch between calorimetric and Cherenkov results, 
was recently restored by the Soudan II calorimeter. The Soudan II
Collaboration reported 
\cite{soudan} a ratio R=(0.64$\pm$0.11(stat$)^{+0.06}_{-0.05}$(sys)) based on 
an exposure of 3.89 kton-year, the largest for a calorimetric 
experiment. 
The analysis of SuperK was based on 33 Kton-year exposure\cite{skana}. 
They selected
4353 fully contained events and 301 partially contained.
Only single ring events were used up to now. They splitted the events in two 
categories: sub Gev events ( $E_{vis}$$<$1.33 GeV) and multi-GeV
events ($E_{vis}$$>$1.33 GeV). They found 
R=(0.63$\pm$0.03(stat)$\pm$0.05(sys)) for the sub-GeV sample and 
R=(0.65$\pm$0.05(stat)$\pm$0.08(sys)) for the multi-GeV sample. 
These results, based on a very large statistical sample, confirmed
the previous measurements. Nevertheless, the interpretation in terms
of neutrino oscillation yields for $\Delta$$m^{2}$ a value
5$\cdot$1$0^{-4}$e$V^{2}$$<$$\Delta$$m^{2}$$<$6$\cdot$1$0^{-3}$e$V^{2}$,
while the best fit for the previous experiments
gave a larger $\Delta$$m^{2}$, for instance Soudan II found 
$\Delta$$m^{2}$=1.1$\cdot$1$0^{-2}$e$V^{2}$. The most promising way to study 
neutrino oscillation using atmospheric neutrino seems to be the evaluation
of $\nu_{\mu}$ flux as a function of L/E: $\Phi_{\nu_{\mu}}$=
$\Phi_{\nu_{\mu}}$(L/E)\cite{pio}. Fig.\ref{f:sk} shows the 
($\Phi_{\nu_{\mu}}$$)_{data}$/($\Phi_{\nu_{\mu}}$$)_{MC}$
as obtained by SuperK. The agreement between the real data
(circles)
and the Monte Carlo expectation(dotted line), obtained supposing a 
$\nu_\mu$$\rightarrow$$\nu_\tau$ or $\nu_\mu$$\rightarrow$$\nu_s$ 
oscillation with parameters $\Delta$$m^{2}$=2.2$\cdot$1$0^{-3}$e$V^{2}$ and 
si$n^{2}$(2$\theta$)=1., is impressive.
From a technical point of view, the SuperK analysis is 
based on single ring events only, therefore the reconstructed energy
E of the event relies only on the muon energy.
The precision in the 
measurement of L is also reduced, since the event kinematic is not completely defined.
The most important goal of the
next generation of atmospheric neutrino detector is the improvement of the L/E ratio
measurement.
Different ideas have been proposed. The measurement of the hadrons produced
in neutrino events with
a highly segmented calorimeter with mass $\simeq$10Ktons or the use of a 
massive calorimeter (M$>$ 30 kton) 
to measure fully contained events with any track multiplicity.
\section{Conclusion}
Usually
the underground detectors for Cosmic Ray studies are much more coarse than 
accelerator experiments.
Such a low granularity comes from the high area/volume
required, imposing an upper limit
on the detector cost and on the number of electronic channels to be taken under
control. Moreover sometimes the coarseness of the experiment is 
fixed by the working conditions.
Nevertheless, a good technical performance is required to these experiments:\\
- {\bf Long data taking.} Underground Observatory looking for Supernova neutrino (LVD,MACRO,
SuperK) have to run for a long time ($>$10 years), avoiding any 
aging effect.\\
- {\bf Flexibility.} An high flexibility is required to these detectors: the physics to be 
investigated spans in energy over more than 3 orders of magnitude. At E$\simeq$10MeV
they look for Supernova neutrino while low energy hadrons produced in 
photonuclear interaction of muon have an energy of few hundreds of MeV.
These detectors measure atmospheric neutrino
with E$\simeq$1 GeV, while the 
average energy of cosmic ray muons detected at Gran Sasso depth
is $E_{\mu}$$\simeq$ 300 GeV. \\
- {\bf Precision measurements.} In some analysis an high rejection factor against
background is required and signals as small as few 1$0^{-4}$ are successfully
measured.\\
About the future the situation is quite different with respect
to the '80s, when the most important lack of underground experiments
was the poor statistics collected. The step forward was the construction of larger
detectors, while now it is not possible to enlarge the underground detectors
area/volume by an order of magnitude. 
Such mass increase is feasible only with underwater Cherenkov detectors.
Moreover the underground space is 
limited and the safety rules became much more restrictive. Special 
care has to be devoted in the choice of the used materials. 
As far as the Cosmic Ray composition study with underground experiments
is concerned, the interpretation of the
results obtained by many experiments is nowadays much more urgent than 
the request for further experimental data. The lack of accelerator measurements 
at high energy and at high rapidity region, introduces large uncertainties 
in the Monte Carlo models for the Cosmic Ray shower development.
The trend for the next generation 
of underground detectors 
seems therefore to be the construction of atmospheric neutrino detectors with
reasonable mass and with an increased granularity.
The ICARUS detector\cite{lvdenergy} for instance is foreseen to run within the 1999 at 
Gran Sasso with 600 tons. It will
take advantage of the superb space resolution to perform an unambiguous
analysis of the $\nu_{\mu}$/$\nu_{e}$ ratio. Other detectors have been
proposed to improve
the L/E measurements for atmospheric neutrino.  
Together with the Long Base Line experiments, 
they will hopefully give an answer on the nature of the neutrino,
using more advanced technology.
\section{Acknowledgments}
I would like to thank G.Battistoni,F.Cei,I.De Mitri,I.Katsavounidis,
S.Kyriazopoulou and O.Palamara for helpful discussions. I am grateful 
to R. Antolini and E.Fantozzi for providing the material for this talk. 

\end{document}